%% file: main.tex
\documentclass[AMA,STIX1COL]{WileyNJD-v2}
\usepackage{float}

\articletype{Research Article}%

\received{26 April 2016}
\revised{6 June 2016}
\accepted{6 June 2016}

\usepackage{blindtext}
\usepackage{subfiles} 
\usepackage{comment}
\begin{document}
\title{Enrollment Forecast for Clinical Trials at the Portfolio Planning Phase Based on Site-Level Historical Data}

\author[1]{Sheng Zhong}

\author[1]{Yunzhao Xing}

\author[1]{Mengjia Yu}

\author[1]{Li Wang}

\authormark{Zhong \textsc{et al}}

\address{\orgdiv{Statistical Innovation Group, Data and Statistical Sciences}, \orgname{AbbVie Inc.}, \orgaddress{\state{Illinois}, \country{United States}}}

\corres{Li Wang,
\email{wangleelee@gmail.com}\\
Sheng Zhong, \email{zhongever@gmail.com}
}

\presentaddress{1 N Waukegan Rd, North Chicago, IL, 60064}

\abstract[Summary]{
Accurate forecast of a clinical trial enrollment timeline at the planning phase is of great importance to both corporate strategic planning and trial operational excellence. While predictions of key milestones such as last subject first dose date can inform strategic decision-making, detailed predictive insights (e.g., median number of enrolled subjects by month for a country) can facilitate the planning of clinical trial operation activities and promote execution excellence. The naïve approach often calculates an average enrollment rate from historical data and generates an inaccurate prediction based on a linear trend with the average rate. The traditional statistical approach utilizes the simple Poisson-Gamma model that assumes time-invariant site activation rates and it can fail to capture the underlying nonlinear patterns (e.g., up-and-down site activation pattern). We present a novel statistical approach based on generalized linear mixed-effects models and the use of non-homogeneous Poisson processes through Bayesian framework to model the country initiation, site activation and subject enrollment sequentially in a systematic fashion. We validate the performance of our proposed enrollment modeling framework based on a set of pre-selected 25 studies from four therapeutic areas. Our modeling framework shows a substantial improvement in prediction accuracy in comparison to the traditional statistical approach. Furthermore, we show that our modeling and simulation approach calibrates the data variability appropriately and gives correct coverage rates for prediction intervals of various nominal levels. Finally, we demonstrate the use of our approach to generate the predicted enrollment curves through time with confidence bands overlaid.            
}

\maketitle

\section{INTRODUCTION}

\subfile{sections/introduction}

\section{INPUT DATA AND FORECAST WORKFLOW}

\subfile{sections/dataworkflow}

\section{METHODOLOGY DETAILS}

\subfile{sections/methods}

\section{PERFORMANCE EVALUATION}

\subfile{sections/performance}

\section{DISCUSSION}

\subfile{sections/discussion}

\newpage
\bibliography{wileyNJD-AMA}

\clearpage

\end{document}

%% file: sections/introduction.tex
In the portfolio planning stage of clinical development, projected trial enrollment duration and associated costs are crucial feasibility factors that senior management has to consider before deciding whether to fund the trial or not. Therefore, a predictive modeling algorithm capable of providing accurate enough enrollment forecast across all portfolios to facilitate management's decision is highly desirable in any pharmaceutical company. The naive and simplest approach currently utilized in practice rests on the concept of average enrollment rate defined as the mean number of subjects enrolled per site per month for a study, often abbreviated as $psm$, $psm = \frac {\text{number of enrolled subjects}} {\text{number of sites } \times \text{ enrollment time}}$ and the assumption that every site in a trial will have such a constant enrollment rate over time. The projected enrollment duration for a planned trial is simply $\frac{\text{planned sample size}}{\text{number of sites proposed } \times \text{ $psm$}}$, where the $psm$ rate in the denominator can be directly taken from the average rate in historical studies, specified based on expert knowledge or obtained in a mixed approach.  It is a fast and easy way to project enrollment duration based on sample size and the total number of sites proposed for the new trial and no deep statistical thinking or theory is involved behind it. Methodology-wise, this approach is purely empirical, does not provide quantification of uncertainty such as confidence intervals, and replies on strong assumption on the constant enrollment rate for all sites overtime, which is often not met in reality. This approach is usually the first tool implemented within a pharmaceutical company, mostly by a team with less formal statistical training. With unsatisfactory predictive performance from this naive approach, a more statistics modeling oriented team is often consulted. Thus the need and the gap of developing a
more systematic and accurate statistical predictive modeling approach pose both opportunities and challenges.   

In the literature, various statistical approaches to modeling and predicting patient accrual have been proposed. The Heitjan et al.\cite{HEITJAN201526, anisimov2016discussion} review paper provides an excellent systematic literature review summarizing existing patient accrual models. Bagiella and Heitjan\cite{https://doi.org/10.1002/sim.843} used a homogenous Poisson process to model the recruitment. Anisimov and Fedorov\cite{https://doi.org/10.1002/sim.2956} improved the recruitment
model by using a Poisson-Gamma mixture model to handle the variation in recruitment rate across multiple centers.
Such type of methods is basically utilizing random effects models\cite{anisimov2007recruitment, anisimov2011statistical, anisimov2020modern}. It aims to capture the heterogeneity in enrollment
rates across different centers whereas the enrollment pattern within each center is described by a homogeneous
Poisson process with gamma distributed rate. Lan et al.\cite{https://doi.org/10.1002/sim.8036} implemented a time-varying rate function that allows modeling
the time decay trend in recruitment while taking site initiation into consideration. Deng et al.\cite{deng2017bayesian} investigated a Bayesian approach to accrual modeling using a non-homogeneous Poisson process where region-specific accrual is accounted in their framework. Zhang and Long\cite{zhang2010stochastic,zhang2012joint} employed a non-homogeneous Poisson process to model patient accrual where the underlying accrual rates are allowed to change over time. Wang et al.\cite{wang2022real} defined time to endpoint maturation framework and linked the concept to key milestone dates in clinical trials. They proposed a simulation based non-homogeneous Poisson process with a normal kernel enrollment rate which can capture the up-and-down enrollment trend in reality and provided improved prediction performance in both simulated and real study enrollment data. The last few works mentioned above have motivated us to apply a non-homogeneous Poisson process through Bayesian framework to model site activation process with some technical adaptation to account for multiple historical studies as input data.   

However, the modeling frameworks above are mainly proposed to be used in trial monitoring at the execution stage where actual patients have already been enrolled into the trial and the goal is to use the accumulated in-trial data up to a certain time point to predict or re-forecast future enrollment beyond. There are not many published literature for enrollment forecast in the portfolio planning phase yet. To fill the gap, we propose a novel statistical framework based on generalized linear
mixed-effects models (GLMM) and the use of non-homogeneous Poisson processes through Bayesian hierarchical framework to model and predict the country initiation, site activation and subject enrollment sequentially in a systematic fashion, utilizing historical site-level enrollment related data. We randomly selected 25 completed studies across four therapeutic areas to validate the performance of our proposed modeling framework. In particular, we compare the prediction accuracy of our proposed modeling framework vs. a previous forecast system developed in our company based on the traditional statistical methods that fit simple probability distributions. Furthermore, we show that our modeling and simulation approach calibrates the data variability appropriately and gives correct coverage rates for prediction intervals of various nominal levels. Finally, we demonstrate the use of our approach to generate the predicted enrollment curves through time with overlaid confidence bands.

%% file: sections/dataworkflow.tex
\subsection{Enrollment framework}
\begin{figure}[h]
\centering
\caption{Three Sequential Segments of Enrollment Framework}
\label{framework_figure}
\includegraphics[width=16cm]{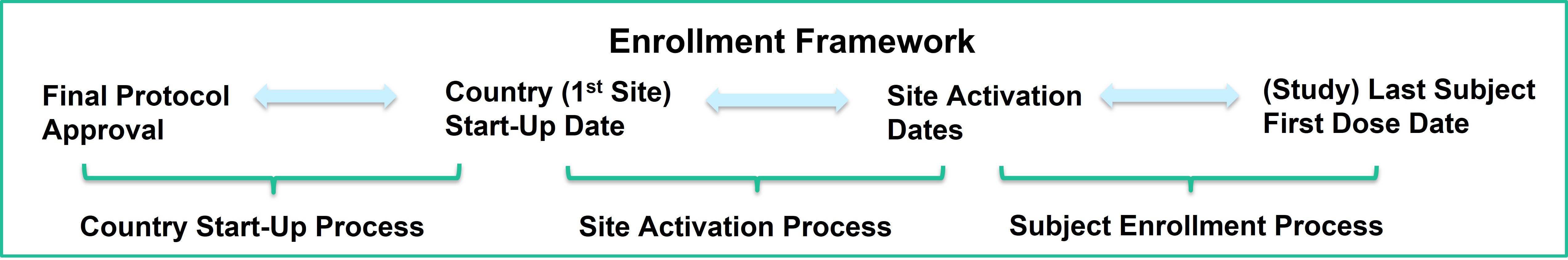}
\end{figure}

The whole enrollment procedure in clinical trials is not just a simple step about subject enrollment. Instead, it is a complicated process with multiple steps. To facilitate the subsequent statistical model development, a comprehensive enrollment framework that clearly defines each step of the whole enrollment procedure needs to be specified first. In the literature, people mainly focused on the subject enrollment process \cite{anisimov2007recruitment, anisimov2011statistical, anisimov2020modern}. However in real-world practice, the time of several important operational steps prior to subject enrollment like country start-up and site start-up would have a crucial impact on the trial duration. In this paper, we dissected the trial duration and established a comprehensive enrollment framework taking into account of all the important operational steps of a clinical trial enrollment. Advanced statistical models are developed (depending on historical data availability) for each segment of the whole enrollment framework. The enrollment framework is defined from the final protocol approval date to the last subject first dose (LSFD) date of a study with three sequential segments as illustrated in Figure \ref{framework_figure}. 

The first segment of the enrollment framework is the country start-up process. The country start-up time is defined as the time from the study start up date to the country start up date. In our consideration, all the selected countries in a clinical trial will share the same study start date which is the final protocol approval date. On the other hand, each country will have its own country start date which is defined in our case as the country's 1st site activation date due to the lack of a specific country start date in our historical data for modeling. The country start-up is a country-level procedure that involves country approval and preparation activities, such as country/region regulatory approval and IRB approval process. It can be affected significantly by a couple of factors, including but not limited to sponsors, countries/regions, therapeutic areas, and phases. For modeling purposes, data availability is also an important factor to be considered, since the final protocol approval date and site activation date are typically not available in the most common public databases. Therefore, dedicated integrated data from multiple data sources are required to estimate the country's start-up time. 

Immediately following the country start-up, the second segment of the enrollment framework is the site activation process. It can be viewed as a stochastic process of site openings starting from the start-up date of the country where the sites locate. The site activation procedure describes a couple of site-level preparation activities for a clinical trial, such as documentation transferring, site training, and drug delivery. It can reflect the relative agility of a site and usually vary among different therapeutic areas, indications, countries, and regions. The site activation date in the definition is the date when the site is ready and opened for patient recruitment in a clinical trial. 

The last part of the enrollment framework is the subject enrollment process. It is a site-specific procedure that starts from the site activation date to the last subject enrolled date of the entire clinical trial. As it's mentioned above, the site starts to recruit patients after the site activation date. In our consideration, the patient enrollment process in each activated site will last until the time of the last patient recruitment of the entire trial. This means that the LSFD date of the entire clinical trial is used as the end of patient enrollment of all sites. Although the last subject enrolled date of each site is also available in our current date set, it is not recommended to be used as the end of the patient enrollment procedure, because once a site is activated, it is capable of recruiting patients until the end of the entire trial enrollment. During the time interval between the last patient enrolled in one particular site and the end of patient enrollment of the entire trial, that particular site is still under the patient enrollment procedure (although not able to recruit any more patients). If that interval is not taken into account, it will introduce extra bias to the length of patient enrollment duration in each site.

\subsection{Historical data}
Two major historical data sources are utilized to model the duration of different segments in the enrollment framework: the site level data from the internal clinical trial management system (CTMS) and the site level data from proprietary Data Query System (DQS) from IQVIA Inc, which contains multiple sponsors' enrollment data. Although there are different limitations in each data source, the combination of these two data sources provides the most comprehensive site level enrollment information in each segment enabling the modeling of the proposed enrollment framework.

In CTMS data, detailed site level timeline from each clinical trial are available. It provides the required anchor dates and process information in the enrollment framework, in particular, the final protocol approval date of each trial. However, the data is limited to the historical clinical trials conducted by AbbVie. It would be challenging to use it alone to plan scenarios where there is no prior AbbVie experience in the indication or country/site.    

On the other hand, in the DQS system, 17 major pharmaceutical companies including AbbVie have signed up to share the operational data from their clinical trials with each other. The data from various sponsors are transferred and integrated into one database. Although the specific names and definitions of fields may vary in the sponsors' internal systems, they are well-aligned and mapped into one standard set of field names and definitions before being transferred into DQS. Moreover, DQS utilized a well-designed algorithm to identify a clinical site across clinical trials and sponsors and assign a unique identifier to it. It allows overlaying the historical information of the same site from different clinical trials and sponsors.

In the proposed enrollment framework, AbbVie CTMS data is used to model country start-up time. The DQS data that includes both AbbVie internal and external studies are used to model both the site activation process and patient enrollment process.

%% file: sections/methods.tex
\subsection{General consideration}
Our primary objective is to forecast the enrollment timeline of a study at the portfolio planning phase where very limited information on the planned study is available. This means any detailed recruitment planning information (e.g., number of planned sites in each country) is not available and actual in-trial enrollment data (e.g., actual number of activated sites up to a cutoff date) are not observed yet for the study, where in-trial enrollment data can include those on country initiation, site activation and subject enrollment etc. For the enrollment forecast purpose, very basic information on the planned study is provided including the total number of subjects to be enrolled, the total number of planned sites, the patient population, the disease indication and therapeutic area of the study.      

Since the in-trial enrollment data is not available at the planning stage, to achieve our primary objective, it is necessary to utilize the enrollment related information from historical studies, both internal or external, to model the underlying enrollment related processes defined in our enrollment framework. With an estimated model for each segment, the Monte Carlo simulation based approach is applied to simulate and forecast the future enrollment activities for the planned study.

As a secondary objective, the prediction of the list of specific countries that enrolled subjects originate from as well as the number of sites activated in each country are often of great interest to the clinical operation team, in addition to the overall subject enrollment timeline. Such country-level detailed insights can facilitate the planning of clinical trial operation activities and promote execution excellence.

The implication on statistical modeling from the need of the country-level predictive insights is that it is sufficient to model the country-specific effects, although our historical study enrollment data distinguishes sites by their unique site identifiers across different studies. Thus observed sites within a country for a study would be treated indistinguishably as drawn randomly from the pool of sites available in the country. This country-level modeling consideration avoids the unnecessary technical complication of modeling the site-specific effects, which is not essential to addressing either our primary or secondary objective.     

\subsection{Country start-up time model}
Suppose we observe the enrollment related data from $S$ historical studies, where these historical studies are usually selected based on having the same therapeutic area, disease indication and patient sub-population (e.g., pediatric patients) as the current planned study. Other study-level features such as study phase (e.g., Phase 2b) or study start year can be used to select recent studies with potentially similar recruitment characteristics. 

For each historical study $j\in \{1,...,S\}$, $N_j$ countries were observed to be initiated and their country start-up dates are available. The initiated countries from different historical studies form a candidate pool of $C$ unique countries, denoted by $\{1,...,C\}$. The observed historical study data come from a list of $N=\sum_{j\in S}{N_j}$ study-country combinations. For the $i$th study-country combination, the study index mapping $j[i]\in \{1,...,S\}$ gives the study index and the country index mapping $k[i]\in \{1,...,C\}$ gives the country index. We note that this notation is employed because different sets of countries (with overlapping) were initiated for subject enrollment for different historical studies. For a historical clinical trial $j$, let $t_{0,j}$ denote the protocol approval date, which is treated as the starting point for the clinical trial recruitment related activities. We define the country start-up time for a country in a study to be the time duration between the study protocol approval date and the country start-up date. In our situation, the country start-up date is chosen to be the country's first site activation date, because our historical data do not include a specific date for country initiation (before its first site activation date). We provide some discussion on other modeling choices in the final section for the scenario that such a specific date for country initiation is collected. Let $t_i$ denote the observed country start-up date for the $i$th study-country combination and $u_i=(t_i-t_{0,j[i]})/m$ denote the corresponding observed country start-up time, where $m$ is a normalization parameter that provides a desired time unit. Note that the observation $u_i$ is the country start-up time for the country $k[i]$ in the historical study $j[i]$. The country start-up time depends on the effects of several factors: country index (e.g., United States), therapeutic area of the study and maybe other covariates. We take a linear mixed-model approach\cite{davison2003statistical, mcculloch2004generalized, mccullagh2019generalized} to modeling the logarithm of the country start-up time,
\begin{equation}\label{csu_mod1}
\mbox{log}(u_i) = \mu_{csu} + \gamma_{csu, k[i]} + \alpha_{csu} X_i + \epsilon_{i}.
\end{equation}
$\mu_{csu}$ is the grand intercept for the model. $\gamma_{csu, k[i]}\sim N(0, \sigma^2_{csu, \gamma})$ denotes the country effect on country start-up times collected from different countries. $\gamma_{csu, k[i]}$ is treated as a random effect, because it is common to enroll subjects from tens of countries for a study. $X_i$ denotes the therapeutic area for the $i$th study-country combination and since only four therapeutic areas (i.e., immunology, oncology, neurology and general medicine) are under consideration in our scenario, $X_i$ is treated as a fixed factor. As $X_i$ has 4 levels, the model matrix for $X_i$ has 3 columns (since the intercept is included) and $\alpha_{csu}$ is a 3 dimensional vector of fixed effect parameters. Finally, $\epsilon_i\sim N(0, \sigma_\epsilon^2)$ denotes the independently and identically distributed normal random error. Other covariates may be added to the model depending on their availability in both the historical data set and at the time of portfolio planning.

If it is believed that country start-up times vary greatly across historical studies, then a random study effect $\beta_{csu, j[i]}\sim N(0, \sigma^2_{csu, \beta})$ (independent of $\gamma_{csu, k[i]}$'s) can be included, as it is common to observe tens of historical studies available from the internal and external databases for modeling purpose. So the country start-up time model can be,
\begin{equation}\label{csu_mod1}
\mbox{log}(u_i) = \mu_{csu} + \beta_{csu, j[i]} + \gamma_{csu, k[i]} + \alpha_{csu} X_i + \epsilon_{i}.
\end{equation}

Furthermore, if exploratory analysis by graphic plots suggests country-dependent therapeutic area effects, then random country slopes can be included with the following model,  
\begin{equation}
\mbox{log}(u_i) = \mu_{csu} + \beta_{csu,j[i]} + \gamma_{csu,k[i]} + (\alpha_{csu} + \delta_{k[i]})X_i + \epsilon_i,
\end{equation}
where $\delta_{k[i]}\sim N(0, \sigma_\delta^2)$ denotes a random effect in the slope of $X_i$.

While the linear mixed-effects model of the logarithm of the country start-up time benefits from its fast computational time, this does not limit us to consider the generalized linear mixed-effects model\cite{mcculloch2004generalized,fitzmaurice2008longitudinal, goldstein2002partitioning} to directly model the outcome variable. For instance, we can model the country start-up time with a Gamma distribution based on the generalized linear mixed-effects model with the log link function,
\begin{equation}
u_i \sim \mbox{Gamma}(\mu_i, \phi) \mbox{ with the Gamma distribution mean } \mu_i = \mbox{exp}(\mu_{csu} + \beta_{csu,j[i]} + \gamma_{csu,k[i]} + \alpha_{csu} X_i),
\end{equation}
where $\phi$ is the dispersion parameter for the Gamma distribution.

For mixed-model estimation, the R package lme4\cite{lme4paper} can be used for estimation of both linear mixed-effects models and generalized linear mixed-effects models. In our case, the study and country effects are modeled as crossed random effects and lme4 package provides a simple syntax and efficient implementation for fitting models with crossed random effects. 

\subsection{Site initiation model}
In our problem, we observe the activation dates for all activated sites within each country in a historical study. So the Bayesian hierarchical non-homogeneous Poisson process approach can be utilized to model the time-varying site activation pattern within a particular country. For instance, in Wang et al.\cite{wang2022real}, this approach has been used to model the subject enrollment process within a specific site based on the observed in-trial data up to a certain time point. In their application, the input data for model estimation is based on the observed data in a single study. We extend this approach to account for multiple historical studies from which recruitment related data are observed. For the $i$th study-country combination defined in the last section, $j[i]$ and $k[i]$ are the corresponding study index and the country index respectively for the $i$th combination. $u_i$, the country start-up time for the $i$th combination, is the starting point of site activation for the $i$th combination. Let $t'_i$ be the end date of site activation for the $i$th combination, where $t'_i$ can be chosen to be the activation date of the last site activated at either the study-country or study level. While the use of the study-country level site activation end date is based on the assumption that the site activation period is country-specific, the use of the study-level site activation end date replies on the assumption that the site activation of all countries would not stop until the target number of sites to be activated has reached. To accommodate the study starting point and the desired time scale, we standardize $t'_i$ and get $u'_i=(t'_i-t_{0,j[i]})/m$ just as we did for $u_i$. Let $N_{in,i}(u)$ be the number of sites activated in the time interval $(u_i,u]$ for $i$th study-country combination. Note that the first activated site is not included in $N_{in,i}(u)$. Then the number of sites activated in $(u_1, u_2]$ for the $i$th study-country combination can be modeled by a Poisson process as follows,
\begin{equation}\label{nh_pp}
      N_{in,i}\left(u_2\right)-N_{in,i}\left(u_1\right) \sim Poisson\left(\int_{u_1}^{u_2}\lambda_{in,i}(v)dv\right), u_2 > u_1 >= u_{i}.
\end{equation}
If it is believed that the site initiation rate starts from a high value and then follows a downward  pattern, the following time-decay rate function can be used:
\begin{equation}
        \lambda^{(1)}_{in,i}(u) = \Lambda_{k[i]}\exp(-(\eta+\epsilon)(u-u_i)), \Lambda_{k[i]} \sim \Gamma(\alpha, \beta), \eta > 0, \epsilon=10^{-5} (\mbox{offset}), u>=u_i.
\end{equation}
The base site activation rate $\Lambda_{k[i]}$ is country-specific and drawn from a Gamma distribution. The posterior predictions of country-specific site activation rates would provide meaningful information to the clinical operation team for their site planning and optimization in different countries. If the exploratory analysis (e.g., times series plotting of site activation dates within countries) shows that the site activation rate generally first increases to a peak and then decays, then the following quadratic rate function would be a good choice,
\begin{equation}
    \lambda^{(2)}_{in,i}(u) = \Lambda_{k[i]}\exp\left(-\frac{(u-u_i-e)^2}{2(\eta+\epsilon)}\right), \Lambda_{k[i]} \sim \Gamma(\alpha, \beta), e>0, \eta >= 0, \epsilon=10^{-5} (\mbox{offset}), u>=u_i    
\end{equation}
The quadratic rate function $\lambda^{(2)}_{in,i}(u)$ is more flexible than the time-decay rate function $\lambda^{(1)}_{in,i}(u)$ because $\lambda^{(2)}_{in,i}(u)$ can absorb the decaying rate function as a special case when the parameter $e$ is close to 0.

For the $i$th study-country combination, the observed data regarding site activation in the time interval $(u_i, u'_i]$ include the total number of sites activated, denoted by $n_{in,i}\left(u'_i\right)$ and the corresponding site activation times for the $i$th combination, i.e., $\overrightarrow{u_{in,i}}=[u_{in,i,m}|m\in\{1,...,n_{in,i}\left(u'_i\right)\}]$. It can be shown that the likelihood function of site activation data over all $N$ study-country combinations with time-decay rate function has the following form,
\begin{equation}
    \begin{split}
        & L\left(\alpha, \beta, \eta;\overrightarrow{u_{in,i}}, n_{in,i}\left(u'_i\right), i\in N\right) = \prod_{i=1}^{N} L\left(\alpha, \beta, \eta;\overrightarrow{u_{in,i}}, n_{in,i}\left(u'_i\right)\right) \propto \\
        & \prod_{i=1}^{N}\left(\frac{\Gamma\left(n_{in,i}\left(u'_i\right)+\alpha\right)}{\Gamma\left(\alpha\right)\beta^{\alpha}}\left(\frac{\beta(\eta+\epsilon)}{\beta\left(1-\exp\left(-\left(\eta+\epsilon\right)\left(u'_i-u_i\right)\right)\right)+(\eta+\epsilon)}\right)^{\left(n_{in,i}\left(u'_i\right)+\alpha\right)} \; \prod_{m }\exp\left(-\left(\eta+\epsilon\right) \left(u_{in,i,m}-u_i\right)\right)\right)
    \end{split}    
\end{equation}
Let $\pi\left(\alpha\right)$, $\pi\left(\beta\right)$, and $\pi\left(\eta\right)$ be the prior distributions for the parameters $\alpha$, $\beta$, and $\eta$ respectively. Then the posterior distribution is,
\begin{equation}\label{pos_dist}
        \pi\left(\alpha, \beta, \eta|\overrightarrow{u_{in,i}}, n_{in,i}\left(u'_i\right), i\in N\right) \propto L\left(\alpha, \beta, \eta;\overrightarrow{u_{in,i}}, n_{in,i}\left(u'_i\right), i\in N\right)\pi\left(\alpha\right)\pi\left(\beta\right)\pi\left(\eta\right)
\end{equation}
By conditioning on $\alpha$, $\beta$, and $\eta$ and the observed data, we can generate $\Lambda_k (k\in \{1,...,C\})$ for country $k$ from the following gamma distribution,
\begin{equation}\label{lambda_cond}
    f\left(\Lambda_k|\alpha, \beta, \eta;\overrightarrow{u_{in,i}}, n_{in,i}\left(u'_i\right), i\in N\right) \sim Gamma\left(\sum_{i:k[i]=k} n_{in,i}\left(u'_i\right) +\alpha, \frac{\beta(\eta+\epsilon)}{\beta\left(1-\exp\left(-\left(\eta+\epsilon\right)\sum_{i:k[i]=k}\left(u'_i-u_i\right)\right)\right)+\left(\eta+\epsilon\right)}\right)
\end{equation}
The likelihood function based on the quadratic rate function and the corresponding posterior distribution are,
\begin{equation}
    \begin{split}
        & L\left(\alpha, \beta, \eta, e; \overrightarrow{u_{in,i}}, n_{in,i}\left(u'_i\right), i\in N\right) = \\ 
		& \prod_{i=1}^N\left(\frac{\Gamma\left(n_{in,i}\left(u'_i\right)+\alpha\right)}{\Gamma\left(\alpha\right)\beta^{\alpha}}\left(\frac{\beta}{\beta\left(\sqrt{2\pi(\eta+\epsilon)}\left[\Phi(\frac{u'_i-u_i-e}{\sqrt{\eta+\epsilon}}) - \Phi(\frac{-e}{\sqrt{\eta+\epsilon}})\right]\right)+1}\right)^{(n_{in,i}\left(u'_i\right)+\alpha)}\prod_{m }\exp\left(-\frac{\left(u_{in,i,m}-u_i-e\right)^2}{2\left(\eta+\epsilon\right)}\right)\right),\\
    \end{split}
\end{equation}
\\
\begin{equation}\label{pos_dist2}
    \pi\left(\alpha, \beta, \eta, e|\overrightarrow{u_{in,i}}, n_{in,i}\left(u'_i\right), i\in N\right) \propto L\left(\alpha, \beta, \eta, e; \overrightarrow{u_{in,i}}, n_{in,i}\left(u'_i\right), i\in N\right)\pi\left(\alpha\right)\pi\left(\beta\right)\pi\left(\eta\right)\pi\left(e\right),
\end{equation}
where $\Phi\left( \cdot \right)$ is the cdf of the standard normal distribution. By conditioning on $\alpha$, $\beta$, $\eta$, $e$ and observed data, we can generate $\Lambda_k$ for country $k$ from the following gamma distribution,
\begin{equation}\label{lambda_cond2}
    \begin{split}
        & f\left(\Lambda_k |\alpha, \beta, \eta, e, \overrightarrow{u_{in,i}}, n_{in,i}\left(u'_i\right), i\in N\right) \sim Gamma\left(\sum_{i:k[i]=k} n_{in,i}\left(u'_i\right)+\alpha, \frac{\beta}{\beta\left(\sqrt{2\pi(\eta+\epsilon)}\sum_{i:k[i]=k}\left[\Phi\left(\frac{u'_i-u_i-e}{\sqrt{\eta+\epsilon}}\right) - \Phi\left(\frac{-e}{\sqrt{\eta+\epsilon}}\right)\right]\right)+1}\right).
    \end{split}
\end{equation}

In the modeling of the rate function for site initiation, we note that a study effect $\Omega_{j[i]}$ may be considered. Unlike the country effect, the study effect is not of direct interest for the simulation of new trials, because a new simulated trial is always a different trial from the historical ones while a common set of countries are used in both historical and future studies. Due to much higher computation complexity from adding an additional study effect and its marginal benefit, we choose to not include it in site initiation modeling.    

For the computation of posterior distribution of model parameters, we use the R function MCMCmetrop1R in the R package MCMCpack\cite{mcmcpackpaper}, which can produce samples from the user derived posterior distribution function (as in \ref{pos_dist} and \ref{pos_dist2}).

\subsection{Subject enrollment model}
In terms of modeling subject enrollment process, our observed data only include site-level summary of subject enrollment information from historical studies, where for each site in a historical study, we observe the total number of enrolled subjects only instead of the actual date when each recruited subject was enrolled. This data limitation does not allow for the modeling of the time dynamics of subject enrollment stochastic process within each site and restricts our approaches to those that model the overall site-level enrollment rates. 

Our historical study data set used for modeling enrollment consists of a number of $N$ study-country combinations from $S$ different studies, where for each study-country combination $i$, the number of sites activated is $n_{in,i}$. Let $N_{en,i,m}$ be the total number of subjects enrolled at the site $m$ in the $i$th study-country combination. Then $N_{en,i,m}$ can be modeled by the following generalized mixed-effects Poisson regression model\cite{mcculloch2004generalized,mccullagh2019generalized, austin2018measures},
\begin{equation}
N_{en,i,m} \sim \mbox{independently as } Poisson(\mu_{im}) \mbox{ where } \log(\mu_{im}) = \mu_{en} + \log(d_{im}) + \gamma_{en, k[i]}.
\end{equation}
The parameter $\mu_{im}$ is the mean number of enrolled subjects for site $m$ in the $i$th study-country combination. $\mu_{en}$ is the grand intercept for the model. The term $\log(d_{im})$ is the offset that accounts for the enrollment duration for the site $m$ in the $i$th study-country combination, where $d_{im}$ is the corresponding total enrollment duration. Here the enrollment duration for a site is defined to be the time duration between the site activation date and the last subject enrollment date that is standardized by the normalization factor $m$. The random parameter $\gamma_{en, k[i]}\sim N(0, \sigma^2_{en,\gamma})$ denotes the country effect and $k[i]$ is the country index for the $i$th study-country combination. If exploratory graphical analysis suggests that the average enrollment rate over sites in a country changes greatly across different historical studies, then the random study effect $\beta_{en, j[i]}$ can be added to the model.  

It is interesting to note that when the parameter $\mu_{en}$ is absorbed in $\gamma_{en,k[i]}\sim N(\mu_{en}, \sigma^2_{en,\gamma})$, then we would essentially assume a lognormal distribution for the country effect $\lambda_{en,k[i]}=\exp(\gamma_{en,k[i]})$ where $N_{en,i,m} \sim Poisson(d_{im}\lambda_{en,k[i]})$. This sheds light on the connection of our proposed model to the popular approach of the Poisson-Gamma model in which, we would impose a Gamma distribution on the parameter $\lambda_{en,k[i]}$.

For the computation, the glmer function in the lme4 package\cite{lme4paper} is used to estimate the mixed-effect poisson regression model with an offset included.

\subsection{Monte Carlo simulation to predict future recruitment process}\label{mc_sim}
The parameters of three enrollment related models (i.e., country start-up time model, site initiation model, and subject enrollment model) need to be estimated before we can proceed to predict future recruitment timeline for the current study under planning. We take the Monte Carlo simulation based approach to randomly generate a number of simulated trials, where the parameters of three recruitment related models are simulated first for each trial and then the country start-up times, site initiation times within each simulated country, and subject enrollment times within each simulated site from each simulated country are simulated sequentially, up to a pre-specified upper time limit, for each simulated trial. We choose 1000 to be the total number of trials to be simulated. The upper time limit can be chosen to be the maximum enrollment duration of all completed historical trials within an organization and it can be therapeutic area dependent. The computation complexity is highly dependent on this upper time limit. Hence we suggest that it is chosen with the knowledge from expert users in the therapeutic area which the current study belongs to, especially when a batch of enrollment scenarios need to be explored. In our case we set the upper time limit to be 5 years. Once all recruitment related data are simulated for all 1000 simulated trials, the median, lower/upper percentiles, and other statistics can be summarized for any quantities of interest. For instance, to provide the study enrollment completion prediction, we can report the median last subject first dose date and its lower and upper percentiles as a prediction interval. The following paragraphs describe in details how to simulate from each of the country start-up time, site initiation and subject enrollment models.

To simulate country start-up times in a new trial, we need to simulate values for the fixed effect parameters $\mu_{csu}$ and $\alpha_{csu}$ and the random effects $\gamma_{csu, k}$ and $\beta_{csu, j}$ as well as the random error $\epsilon_i$. For the fixed parameters $\mu_{csu}$ and $\alpha_{csu}$, we draw from the asymptotic normal distribution of their restricted maximum likelihood (REML) estimators, where the mean of the normal distribution equals the REML estimates and the standard deviation equals the standard error of REML estimates. For the random country effect $\gamma_{csu, k}$ for a country $k\in \{1,...,C\}$, we simulate from a normal distribution with mean and variance equal to the conditional mode and conditional variance respectively of the random effect $\gamma_{csu, k}$, because the countries we use in the new study are the ones used in the historical studies. For the random study effect $\beta_{csu, j}$, we simulate from its unconditional distribution $N(0, \hat{\sigma}^2_{csu, \beta})$ where $\hat{\sigma}^2_{csu, \beta}$ is the REML estimate of $\sigma^2_{csu, \beta}$, because a study under planning is a new study different from the historical ones. For the random error $\epsilon_i$, we simulate from the normal distribution $N(0, \hat{\sigma}^2_\epsilon)$ where $\hat{\sigma}^2_\epsilon$ is the REML estimate of $\sigma^2_\epsilon$. Once all the parameters are simulated, they can be combined by, for instance, the equation \ref{csu_mod1}, to produce the country start-up time for country $k$ in a simulated trial $j$.

To conduct site initiation simulation, we need to first draw the values for model parameters (i.e., $\alpha$, $\beta$, $\eta$ and $e$) from the MCMC samples for the posterior distribution of either \ref{pos_dist} or \ref{pos_dist2}. The country-specific base site activation rate $\Lambda_{k}$ is drawn from the Gamma distribution of either \ref{lambda_cond} or \ref{lambda_cond2}, conditional on the previously drawn model parameter values and observed historical data. Then the total number of sites to be activated and the site activation times can be drawn following the non-homogeneous Poisson process defined by \ref{nh_pp}.  

To conduct subject enrollment simulation, we follow the similar approach as in country start-up time simulation to simulate values for the fixed effect parameter $\mu_{en}$ and the random effects $\gamma_{en, k}$ and $\beta_{en, j}$ for a country $k$ in a simulated trial $j$. The enrollment duration offset is set according to the pre-specified upper time limit for simulation, 5 years. Then the total number of subjects to be enrolled and the subject enrollment times can be drawn following the homogeneous Poisson process.

%% file: sections/performance.tex
\subsection{General consideration}
To validate the performance of our proposed modeling framework, we select a set of recently completed studies within our organization, run our models to generate the predictions on enrollment duration, and compare the predictions to the ground-truth values. The enrollment duration of a validation study is defined to be the time duration between the protocol approval date and the last subject first dose date. To get the enrollment prediction for a validation study, the inputs provided to our model include the total number of subjects, the total number of sites, the therapeutic area, disease indication and the patient population of the validation study. In addition, the enrollment related data from historical studies with the same disease indication and patient population as the validation study are used for estimation of model components. Here historical studies are those studies that have their enrollment completed prior to the protocol approval date of the validation study. We compare the performance of our proposed modeling framework to a previous internally-developed enrollment forecast system based on the traditional statistical methods that fit simple probability distributions, described in the next subsection. Furthermore, we show that our modeling and simulation framework calibrates the data variability correctly by comparing the nominal level vs. the coverage rate for prediction intervals of various nominal levels. Finally, we demonstrate how to generate the predicted enrollment curves through time, overlaid with confidence bands, which are deemed very informative to trial operation planning by our users.  

\subsection{A previous enrollment forecast system}
In a previous enrollment forecast system developed internally, the whole enrollment procedure is also divided into three segments with different definitions. The first segment is called site start-up period-1, which is the period from the final protocol approval date to the open date of each site. The second segment is site start-up period-2, which is defined as the time duration between the site open date and the first subject enrollment date for the site. The last segment is the enrollment period  which is from the first subject enrollment date for the site to the last subject enrollment date of the whole study. The models underlying the previous forecast system are primitive compared to our proposed modeling framework. For site start-up period-1, no formal statistical modeling is employed and the simulation is simply based on bootstrapping the observed periods from the historical data in the same therapeutic area and country with replacement. For the site start-up period-2, the modeling utilizes an log-normal distribution to fit historical data for each country and then random samples are drawn from it for various sites in each country. For the modeling of subject enrollment, a country-level modeling approach is taken. A separate Gamma distribution is fitted to the historically observed enrollment rates of all sites in each country and then during the simulation stage, the enrollment rate for each site is randomly drawn from the Gamma distribution and used as the parameter of Poisson distribution to simulate subject enrollment in each day. 

\subsection{Selection of evaluation studies}
To ensure the objectivity of the study selection process, an independent steering committee other than the modeling team (which the authors belong to) forms to lead the selection of a set of enrollment completed studies for model evaluation. First, a candidate set of studies are pulled from our internal database, based on a list of search criteria given in Table \ref{study_sel_crit}. Then for each candidate study, the actual enrollment curve is plotted for the purpose of checking unusual shapes, such as pauses and drastic enrollment speed changes. These unusual shapes are often due to rare unforeseeable events. Whenever such events happen, our internal users would manually assess the impact on a case-by-case basis and we shall not expect any models can still provide accurate predictions. Based on the manual inspection of the enrollment curves, a list of studies with unusual shapes in enrollment curves is proposed so as to be removed from model performance evaluation, and the steering committee makes the final decision on whether a study should be included in the list of evaluation studies. Ultimately, a total of 25 studies are selected from four different therapeutic areas to test the performance of our proposed modeling approach in comparison to the previous internally-developed forecast system. Table \ref{study_ta_count} shows the break-up of 25 studies into four therapeutic areas.

\begin{table}
\caption{Selection criteria for a candidate set of studies for model performance validation}
\label{study_sel_crit}
\begin{center}
\begin{tabular}{c|c}
\hline
Criteria & Value \\ \hline
Enrollment completion date & Within past 3.5 years\\
Interventional vs. observational studies & Interventional studies only  \\
Study phase & 2 and 3\\
Therapeutic area & Oncology, neuroscience, immunology, and general medicine\\
Total number of sites & 10-500 sites\\
Total number of subjects & 50-1600 subjects\\
\hline
\end{tabular}
\end{center}
\end{table}

\begin{table}
\caption{Number of selected studies in different therapeutic areas}
\label{study_ta_count}
\begin{center}
\begin{tabular}{c|cccc}
\hline
 & Immunology & General Medicine & Neuroscience & Oncology   \\ \hline
Number of studies  &  14 & 4 & 3 & 4\\
\hline
\end{tabular}
\end{center}
\end{table}

\subsection{Prediction accuracy}
To evaluate the model performance, we apply both our proposed modeling framework and the previous enrollment forecast system to the 25 studies selected by the independent steering committee. In terms of the performance evaluation metrics, we report the rates of coverage of the ground-truth enrollment duration by prediction intervals with fixed radii set to be +/- 1, 2, and 3 months (so the widths of prediction intervals are 2, 4 and 6 months respectively) as well as the coverage rate by the prediction interval of non-fixed width with nominal confidence level 95\%. In addition, we also look at the mean and median absolute prediction error of the median predicted enrollment duration compared to the ground-truth.

From Table \ref{perform_metrics}, our proposed approach achieves higher coverage rates for all prediction intervals of various fixed widths than the previous enrollment forecast system. In terms of median absolute prediction error, our proposed approach achieves 50\% reduction from 5.6 months to 2.8 months, while for the mean absolute prediction error that is often influenced by large errors, our method still provides 17\% reduction from 5.8 months to 4.8 months. This suggests that the traditional distribution-based approach is not adequate to capture the the sophisticated patterns behind recruitment related processes, compared to our more complex modeling approach based on mixed-effects models and non-homogeneous Poisson processes models. Table \ref{pred_25_studies_new} and \ref{pred_25_studies_old} provide the detailed outputs from our new modeling approach and the previous forecast system respectively.

\begin{table}
\caption{Predictions of enrollment duration for 25 studies by our new modeling framework vs. a previous internally-developed enrollment forecast system}
\label{perform_metrics}
\begin{center}
\begin{tabular}{c|ccccccc}
\hline
 & Coverage of & Coverage of & Coverage of & Coverage of & Width of & Median Absolute & Mean Absolute\\
Model & 2-Month PI & 4-Month PI  & 6-Month PI & 95\% PI & 95\% PI & Prediction Error & Prediction Error\\ \hline
New & 7 (28\%) & 8 (32\%) & 13 (52\%) & 23 (92\%) & 20 months & 2.8 months & 4.8 months\\
Previous & 5 (20\%) & 6 (24\%) & 9 (36\%) & 5 (20\%)	& 2 months & 5.6 months & 5.8 months\\
\hline
\end{tabular}
\end{center}
\end{table}

\begin{table}
\caption{Nominal level vs. actual coverage rate of prediction intervals from our new modeling framework}
\label{coverage_rates}
\begin{center}
\begin{tabular}{c|cccc}
\hline
Nominal Level of PI &	40\% & 50\% & 70\% & 95\% \\ \hline
Number of studies correctly predicted & 13  & 15  & 17 & 23 \\
Coverage rate &	52\% & 60\% & 68\% & 92\%\\
SE for coverage rate	& 10\% & 10\% & 9\% & 5\% \\
95\% CI for coverage rate & 32\%-72\%	& 40\%-80\%	& 50\%-86\%	& 82\%-100\% \\
Prediction interval width & 5.3 Months & 7 Months & 10 Months & 20 months \\
\hline
\end{tabular}
\end{center}
\end{table}

\begin{table}
\caption{Predictions of enrollment duration for 25 studies by our new modeling framework (unit:month) }
\label{pred_25_studies_new}
\begin{center}
\begin{tabular}{c|cccccc}
\hline
Study & Actual & Median Predicted  &  & Within & Within & Within \\
NO. & Duration &  Duration (95\% PI) & Residual & +/- 1 Month & +/- 2 Months & +/- 3 Months \\\hline
1	&	21.23	&	23.94	(	14.66	,	44.90	)	&	2.70	&	No	&	No	&	Yes	\\
2	&	20.93	&	20.26	(	14.02	,	34.03	)	&	-0.67	&	Yes	&	Yes	&	Yes	\\
3	&	21.87	&	15.17	(	10.38	,	22.68	)	&	-6.69	&	No	&	No	&	No	\\
4	&	15.63	&	16.16	(	9.88	,	30.08	)	&	0.53	&	Yes	&	Yes	&	Yes	\\
5	&	22.47	&	29.22	(	11.45	,	Inf	)	&	6.75	&	No	&	No	&	No	\\
6	&	26.37	&	20.46	(	13.71	,	31.27	)	&	-5.90	&	No	&	No	&	No	\\
7	&	23.20	&	23.63	(	17.58	,	34.35	)	&	0.43	&	Yes	&	Yes	&	Yes	\\
8	&	26.83	&	24.31	(	16.07	,	40.46	)	&	-2.53	&	No	&	No	&	Yes	\\
9	&	30.27	&	20.91	(	10.95	,	51.80	)	&	-9.35	&	No	&	No	&	No	\\
10	&	37.33	&	25.31	(	15.37	,	55.11	)	&	-12.02	&	No	&	No	&	No	\\
11	&	28.77	&	26.02	(	17.11	,	40.29	)	&	-2.74	&	No	&	No	&	Yes	\\
12	&	19.93	&	19.18	(	12.29	,	34.41	)	&	-0.75	&	Yes	&	Yes	&	Yes	\\
13	&	17.53	&	20.18	(	13.54	,	35.40	)	&	2.65	&	No	&	No	&	Yes	\\
14	&	11.27	&	10.13	(	6.38	,	17.34	)	&	-1.14	&	No	&	Yes	&	Yes	\\
15	&	10.03	&	10.07	(	6.89	,	17.29	)	&	0.03	&	Yes	&	Yes	&	Yes	\\
16	&	21.13	&	22.11	(	15.72	,	36.70	)	&	0.98	&	Yes	&	Yes	&	Yes	\\
17	&	19.93	&	12.77	(	7.96	,	20.31	)	&	-7.16	&	No	&	No	&	No	\\
18	&	16.77	&	19.55	(	9.96	,	60.24	)	&	2.79	&	No	&	No	&	Yes	\\
19	&	18.37	&	14.86	(	9.24	,	25.40	)	&	-3.50	&	No	&	No	&	No	\\
20	&	30.80	&	19.14	(	12.48	,	34.99	)	&	-11.66	&	No	&	No	&	No	\\
21	&	24.90	&	18.61	(	9.39	,	46.12	)	&	-6.29	&	No	&	No	&	No	\\
22	&	26.37	&	31.81	(	21.82	,	51.40	)	&	5.44	&	No	&	No	&	No	\\
23	&	20.63	&	11.58	(	6.90	,	17.76	)	&	-9.05	&	No	&	No	&	No	\\
24	&	16.10	&	35.03	(	16.21	,	Inf	)	&	18.93	&	No	&	No	&	No	\\
25	&	10.83	&	11.58	(	8.11	,	17.70	)	&	0.75	&	Yes	&	Yes	&	Yes	\\

\hline
\end{tabular}
\end{center}
\end{table}

\begin{table}
\caption{Predictions of enrollment duration for 25 studies by a previous internally-developed enrollment forecast system (unit:month) }
\label{pred_25_studies_old}
\begin{center}
\begin{tabular}{c|cccccc}
\hline
Study & Actual & Median Predicted  &  & Within & Within & Within \\
NO. & Duration &  Duration (95\% PI) & Residual & +/-1 Month & +/-2 Months & +/-3 Months \\\hline
1	&	21.23	&	24.07	(	22.20	,	26.10	)	&	2.83	&	No	&	No	&	Yes	\\
2	&	20.93	&	21.37	(	19.90	,	22.83	)	&	0.43	&	Yes	&	Yes	&	Yes	\\
3	&	21.87	&	13.53	(	12.87	,	14.17	)	&	-8.33	&	No	&	No	&	No	\\
4	&	15.63	&	14.23	(	13.47	,	14.93	)	&	-1.40	&	No	&	Yes	&	Yes	\\
5	&	22.47	&	19.00	(	18.03	,	20.13	)	&	-3.47	&	No	&	No	&	No	\\
6	&	26.37	&	17.67	(	16.87	,	18.37	)	&	-8.70	&	No	&	No	&	No	\\
7	&	23.20	&	23.70	(	22.50	,	24.70	)	&	0.50	&	Yes	&	Yes	&	Yes	\\
8	&	26.83	&	14.33	(	13.47	,	15.20	)	&	-12.50	&	No	&	No	&	No	\\
9	&	30.27	&	26.50	(	24.37	,	28.63	)	&	-3.77	&	No	&	No	&	No	\\
10	&	37.33	&	25.80	(	24.47	,	27.60	)	&	-11.53	&	No	&	No	&	No	\\
11	&	28.77	&	16.30	(	15.73	,	16.83	)	&	-12.47	&	No	&	No	&	No	\\
12	&	19.93	&	20.40	(	19.00	,	22.17	)	&	0.47	&	Yes	&	Yes	&	Yes	\\
13	&	17.53	&	13.50	(	12.67	,	14.23	)	&	-4.03	&	No	&	No	&	No	\\
14	&	11.27	&	9.07	(	8.47	,	9.90	)	&	-2.20	&	No	&	No	&	Yes	\\
15	&	10.03	&	9.60	(	8.83	,	10.40	)	&	-0.43	&	Yes	&	Yes	&	Yes	\\
16	&	21.13	&	15.33	(	14.47	,	16.10	)	&	-5.80	&	No	&	No	&	No	\\
17	&	19.93	&	13.43	(	12.63	,	14.13	)	&	-6.50	&	No	&	No	&	No	\\
18	&	16.77	&	14.40	(	13.37	,	15.43	)	&	-2.37	&	No	&	No	&	Yes	\\
19	&	18.37	&	12.77	(	11.73	,	13.63	)	&	-5.60	&	No	&	No	&	No	\\
20	&	30.80	&	19.90	(	18.83	,	21.23	)	&	-10.90	&	No	&	No	&	No	\\
21	&	24.90	&	32.07	(	29.07	,	36.37	)	&	7.17	&	No	&	No	&	No	\\
22	&	26.37	&	34.73	(	30.27	,	39.97	)	&	8.37	&	No	&	No	&	No	\\
23	&	20.63	&	11.40	(	10.63	,	12.20	)	&	-9.23	&	No	&	No	&	No	\\
24	&	16.10	&	31.07	(	27.37	,	35.00	)	&	14.97	&	No	&	No	&	No	\\
25	&	10.83	&	11.73	(	10.80	,	12.70	)	&	0.90	&	Yes	&	Yes	&	Yes	\\
\hline
\end{tabular}
\end{center}
\end{table}
\subsection{Coverage of prediction intervals}
In term of the prediction interval of nominal level 95\%, our method has achieved the actual coverage rate of 92\% (see Table \ref{perform_metrics}), while the previous forecast system is ill-calibrated and gives a coverage rate of 20\% together with an extremely narrow width. This implies that the variability in historical data is not accounted correctly in either modeling or simulation in the previous enrollment forecast system. While the prediction intervals of other nominal levels are not available in the previous forecast system, we further investigate the discrepancy between the normal level vs. actual coverage rate of our proposed new modeling approach. Table \ref{coverage_rates} shows that the nominal levels are very consistent with the actual coverage rates at different nominal levels. This is because our methodology properly accounts for the data variability at different levels (such as study and country level) with appropriate random effects. In the simulation stage, the random effects are simulated from the correct conditional or unconditional asymptotic sampling distributions correspondingly (see subsection \ref{mc_sim}).  Although 95\% is a widely used number for the level of significance in statistical hypothesis testing, the width of 95\% prediction intervals is in general too large to be used in practice in our scenario. This is often due to the large amount of variability naturally in enrollment related processes and the limited number of historical studies available for model estimation. The 40\% or 50\% prediction intervals tend to offer a good balance between the coverage rate and the interval width in our application.    

\subsection{Predicted enrollment curves}
\begin{figure}[h]
\centering
\caption{Predicted enrollment curve with 95\% confidence bands}
\label{enroll_curve_95}
\includegraphics[width=16cm]{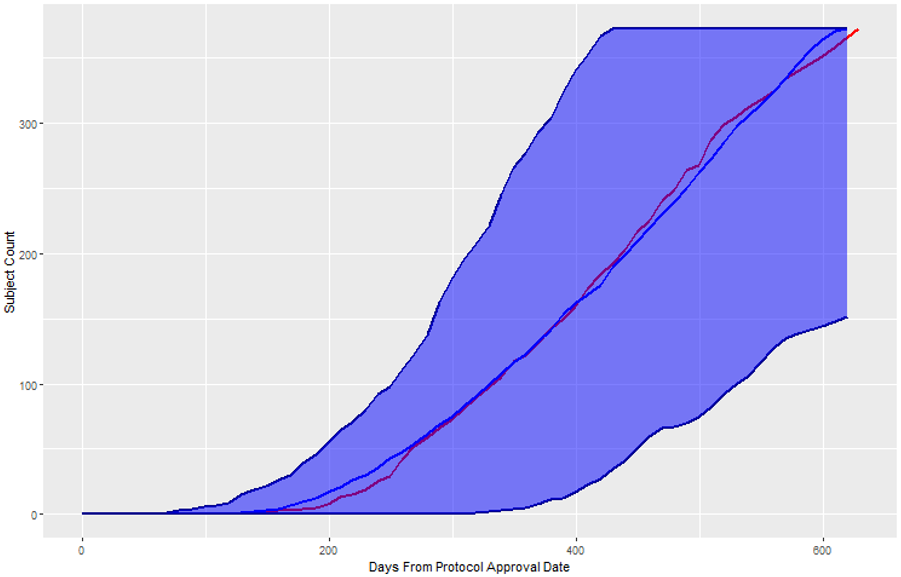}
\end{figure}

\begin{figure}[h]
\centering 
\caption{Predicted enrollment curve with 40\% confidence bands}
\label{enroll_curve_40}
\includegraphics[width=16cm]{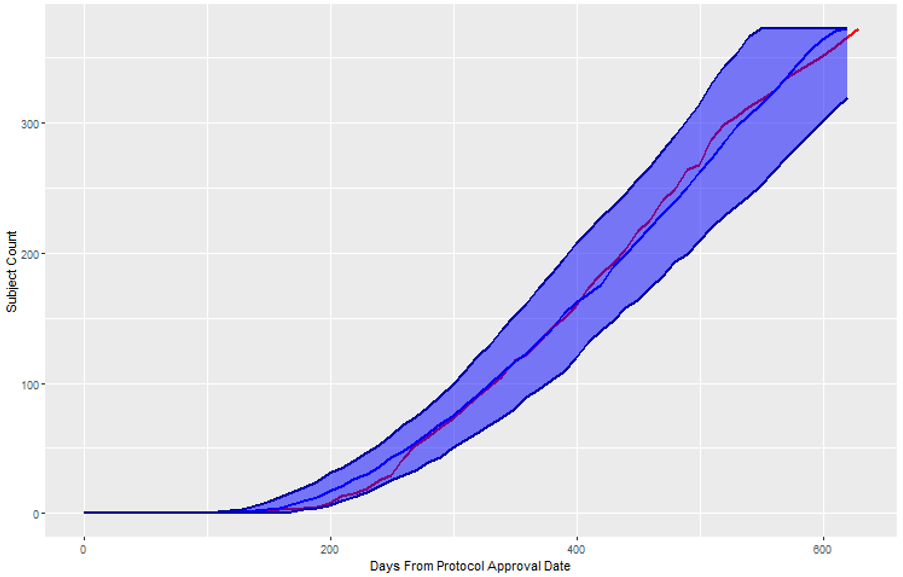}
\end{figure}

In addition to the predicted enrollment duration, our users also find it very informative to their planning job to have a predicted enrollment curve throughout the whole time course from protocol approval to enrollment completion. This can be achieved by first selecting a grid of equally spaced time points (e.g., in months) and then summarizing the median, low percentile, upper percentile total enrollment number for each time point in the grid. In Figure \ref{enroll_curve_95} and \ref{enroll_curve_40}, we show the predicted enrollment curves with 95\% and 50\% confidence bands for a well-predicted study. Note that the upper confidence band is capped by the target number of subjects enrolled for the study, as preferred by our users.  

%% file: sections/discussion.tex
For the problem of forecasting enrollment at the planning stage, we have developed a novel statistical modeling framework, which is based on GLMM and the use of non-homogeneous Poisson processes through Bayesian framework to systematically model the country initiation, site activation and subject enrollment in sequential steps. Our new modeling framework shows a substantial improvement in prediction accuracy in comparison with the traditional statistical approach that fits simple probability distributions, based on a collection of 25 pre-selected studies from four therapeutic areas. Furthermore, we have showed that our modeling and simulation framework calibrates the data variability appropriately and gives correct coverage rates for prediction intervals of various nominal levels. With the Monte Carlo simulation, we demonstrated how to generate the median predicted enrollment curve with confidence bands. It is no harder to generate predictions on other recruitment related quantities such as the median number of sites for each country and the median country start date.

In terms of the modeling of country start-up time, we utilize a generalized linear mixed-effects model to account for the country-level and study-level variability in historical studies as well as the effects of fixed covariates. Machine learning models (e.g., gradient boosting) can be exploited if more covariates are available for modeling. But appropriately accounting for the variability of predictions from machine learning models is not a easy task. Another potential caveat is that more covariate information is needed at the time of enrollment prediction. For a study at the portfolio planning stage, often limited information is known about the study.   

With regard to the modeling of site activation process within a country, if a specific country initiation date is available, we may need to consider the fact that the time duration between the country initiation date and its first site activation date could be much longer than the times between successive activation dates of subsequently activated sites. In this case, we would recommend to model the time between country initiation and its first site activation separately from the rest of the activated sites. An alternative approach is to incorporate the time between country initiation and its first site activation into the definition of country start-up time and handle it implicitly in the modeling of the country start-up times.

For the modeling of subject enrollment, our observed data only include site-level aggregate subject enrollment information from historical studies. If instead the actual date when each subject was enrolled is observed, a non-homogeneous Poisson process can be used to estimate the time-varying rate function for subject enrollment process in a site with each country. Likewise, we may need to take into consideration that it often takes longer time to enroll the first subject due to the overhead site preparation work and hence modeling the time to the first subject enrollment separately may be more appropriate.

When using an enrollment forecast system that depends on historical studies for model estimation, it is important for the end user (e.g., clinical operation team) to conduct data-checking for the historical studies being used. If the summary statistics on, for instance, enrollment speed, in historical studies are systematically slower or faster than what the user expects, caution needs to be taken when interpreting the predicted enrollment duration. Certain ad-hoc measures may be considered, such as excluding some studies with extreme statistics or using a multiplier to adjust the speed of enrollment. 

\subsection*{Disclosure}
This manuscript was sponsored by AbbVie. AbbVie contributed to the design, research, and interpretation of data, writing, reviewing, and approved the content.   All authors are employees of AbbVie Inc. and may own AbbVie stock.

\subsection*{Data availability statement}

The authors elect to not share data.